\documentclass[10pt,conference]{IEEEtran}
\IEEEoverridecommandlockouts
\usepackage{cite}
\usepackage{amsmath,amssymb,amsfonts}
\usepackage{algorithmic}
\usepackage{graphicx}
\usepackage{textcomp}
\usepackage{xcolor}
\usepackage{url}
\usepackage{multirow}
\usepackage{subfigure}
\usepackage{comment}
\usepackage{enumitem}
\IEEEoverridecommandlockouts

\def\BibTeX{{\rm B\kern-.05em{\sc i\kern-.025em b}\kern-.08em
    T\kern-.1667em\lower.7ex\hbox{E}\kern-.125emX}}

\begin{document}

\title{Industrial-Scale Neural Network Clone Detection with Disk-Based Similarity Search}

\author{
\IEEEauthorblockN{Gul Aftab Ahmed}
\IEEEauthorblockA{\textit{Lero Research Centre}\\
\textit{Trinity College Dublin} \\
Dublin, Ireland \\
ahmedga@tcd.ie}
\and
\IEEEauthorblockN{Muslim Chochlov}
\IEEEauthorblockA{\textit{Lero Reseach Centre}\\
\textit{University of Limerick}\\
Limerick, Ireland \\
muslim.chochlov@ul.ie}
\and
\IEEEauthorblockN{Abdul Razzaq
}
\IEEEauthorblockA{\textit{Lero Research Centre} \\
\textit{University of Limerick}\\
Limerick, Ireland \\
abdul.razzaq@ul.ie}
\and
\IEEEauthorblockN{James Vincent Patten}
\IEEEauthorblockA{\textit{Lero Research Centre} \\
\textit{University of Limerick}\\
Limerick, Ireland \\
james.patten@ul.ie}
\and 
\IEEEauthorblockN{Yuanhua Han}
\IEEEauthorblockA{\textit{WN Digital IPD and} \\
\textit{Trustworthiness Enabling} \\
\textit{Huawei Technologies Co., Ltd.}\\
Xi'an, Shaanxi, China \\
hanyuanhua2@huawei.com}
\and
\IEEEauthorblockN{Guoxian Lu}
\IEEEauthorblockA{\textit{WN Digital IPD and} \\
\textit{Trustworthiness Enabling} \\
\textit{Huawei Technologies Co., Ltd.}\\
Shanghai, China \\
luguoxian@huawei.com}

\and
\IEEEauthorblockN{Jim Buckley}
\IEEEauthorblockA{\textit{Lero Research Centre} \\
\textit{University of Limerick}\\
Limerick, Ireland \\
jim.buckley@ul.ie}
\and
\IEEEauthorblockN{David Gregg}
\IEEEauthorblockA{\textit{Lero Research Centre}\\
\textit{Trinity College Dublin} \\
Dublin, Ireland \\
david.gregg@cs.tcd.ie}
}

\hyphenation{DB-SSCD}

\maketitle

\begin{abstract}
Code clones are similar code fragments that often arise from copy-and-paste programming. Neural networks can classify pairs of code fragments as clone/not-clone with high accuracy. However, finding clones in industrial-scale code needs a more scalable approach than pairwise comparison.  We extend existing neural network-based clone detection schemes to handle codebases that far exceed available memory, using indexing and search methods for external storage such as disks and solid-state drives. We generate a high-dimensional vector embedding for each code fragment using a transformer-based neural network. We then find similar embeddings using efficient multidimensional nearest neighbor search algorithms on external storage to find similar embeddings without pairwise comparison.
%We use an efficient multidimensional nearest neighbor search algorithm to find similar high-dimensional embeddings generated from code fragments by a neural network.
% Detecting code clones at an industrial scale requires a scalable approach using 

We identify specific problems with industrial-scale code bases, such as large sets of almost identical code fragments that interact poorly with $k$-nearest neighbour search algorithms,
and provide an effective solution. We demonstrate that our disk-based clone search approach achieves similar clone detection accuracy as an equivalent in-memory technique. Using a solid-state drive as external storage, our approach is around 2$\times$ slower than the in-memory approach for a problem size that can fit within memory. We further demonstrate that our approach can scale to over a billion lines of code, providing valuable insights into the trade-offs between indexing speed, query performance, and storage efficiency for industrial-scale code clone detection.

%We introduce DB-SSCD, a scalable system that offers two disk-based nearest neighbor search options: Faiss with an inverted file index (IVF) and Milvus with a graph-based index (DiskANN).
%% Here we should say something about the performance of our approach in absolute terms, or relative to other clone detection tools.
%On a dataset of 1.16 billion lines of code, we demonstrate that Faiss completes indexing 3.4× faster than Milvus, while Milvus provides 2.04× higher query throughput. Our experiments demonstrate the system’s ability to scale to over a billion lines of code, providing valuable insights into the trade-offs between indexing speed, query performance, and storage efficiency, making DB-SSCD a practical solution for large-scale code clone detection.
\end{abstract}

% \section{Introduction}

\section{Motivation}
Two segments of source code are considered clones if they are identical or similar. Code clones often arise from copying code within or between programs. Maintaining multiple copies of code can be hard, and there are questions about the ownership, vulnerabilities and licence conditions of copied code \cite{Ain2019,Rattan2013}. Clone detection tools attempt to identify code clones, thus helping address some of these issues \cite{ahmed2024nearest}.

Neural networks, particularly transformer-based models like CodeBERT, have proven extremely effective in solving the clone comparison problem, where two code segments are directly compared in a pairwise manner to determine whether they are clones \cite{Wang2021,Feng2020, White2016, Wei2017, Saini2018, Guo2020}.
%\cite{Wang2021,Feng2020, Guo2020}. 
%\cite{Wang2021,Feng2020, White2016, Wei2017, Saini2018, Buch2019, Zhang2019, Guo2020}.
However, there has been much less success in using neural networks to solve the \textit{clone search} problem with millions of lines of code. One notable exception is SSCD \cite{Chochlov2022}, which uses the CodeBERT \cite{Feng2020} neural network to generate a fixed-length vector (or \textit{embedding}) for each code fragment. SSCD then applies efficient $k$-approximate nearest neighbour search ($k$-ANN) to find code segments with similar embeddings. Thus, neural networks can be used to find similar code segments without $O(N^2)$ pairwise comparisons.

While SSCD demonstrates that $k$-ANN can efficiently identify code clones across large codebases, the SSCD approach remains limited by memory constraints, making it less suitable for industrial-scale repositories that exceed available RAM. To address this, we introduce DB-SSCD, a disk (or solid-state drive)-based scalable solution for neural network-based code clone detection. DB-SSCD leverages efficient indexing techniques to scale clone detection to massive, industrial-sized codebases, while maintaining accuracy levels comparable to in-memory methods. By utilizing disk-based nearest neighbor search, DB-SSCD handles codebases significantly larger than what memory-bound systems can accommodate.

To solve the problem at industrial-scale, our solution must achieve accuracy similar to existing tools, while scaling to industrial-sized code bases. The evaluation of our solution is therefore centered around two research questions:
\begin{itemize}
\item \textbf{RQ1: Can our solution achieve clone search accuracy similar to existing scalable clone detection tools?}
\item \textbf{RQ2: Can our solution scale to clone search in industrial-sized code with at least one billion lines of code, and where the memory requirements of the search data structures exceed the size of RAM?}
\end{itemize}

% Gul's suggested contibutions
%With DB-SSCD, we aim to answer two key research questions:
%\begin{itemize}
%    \item RQ1: Does DB-SSCD maintain the recall achieved by SSCD in detecting complex clone types on the standard BigCloneBench dataset?
%    \item (RQ2): What are the performance characteristics of DB-SSCD in terms of indexing and search efficiency, and scalability when applied to codebases exceeding RAM capacity?
%\end{itemize}

\setlist{topsep=0pt, leftmargin=*}

\noindent We evaluate the clone search accuracy of DB-SSCD using the BigCloneBench\cite{Svajlenko2017a} clone detection dataset, which consists of 320 million lines of Java source code. There is no clone detection dataset with more than a billion lines of code. We therefore evaluate the scalability of our techniques on The Stack dataset \cite{Kocetkov2022TheStack} from the BigCode project (bigcode-project.org), specifically the C subset, which contains 1.06 billion lines of code. We make the following contributions:
\begin{itemize}
     \item We propose DB-SSCD, a scalable system for neural network-based clone detection that leverages disk-resident nearest-neighbour indexing and search algorithms. %By extending nearest-neighbor search to disk, we demonstrate that DB-SSCD achieves comparable recall rates to in-memory approaches on the Full BCB dataset (replicating the recall achieved by Chochlov et al. \cite{Chochlov2022}), showing that it retains high accuracy while scaling to larger datasets.
     \item We propose a de-duplication method that replaces Type-1 (identical) clone classes with a single representative embedding, enabling $k$-ANN algorithms to be more effective by searching over fewer neighbors, $k$.
     %\item We demonstrate the effectiveness of our solution with two disk-based nearest-neighbour search systems: Faiss (Disk-IVF) and Milvus (DiskANN).
     \item We demonstrate DB-SSCD clone search is as accurate as equivalent in-memory techniques, although about $2\times$ slower. %DB-SSCD is much more accurate than SourcererCC on the BCB-Full dataset, and around $3\times$ faster.

     \item We validate the scalability of DB-SSCD on real-world, industrial-sized codebases, processing over 1 billion lines of code.% and 18.2 million functions from the BigCode Stack C dataset. This demonstrates that disk-based indexing systems can handle datasets that far exceed available memory, making neural network-based clone detection feasible at an industrial scale.
    
    %\item We evaluate two disk-based search systems: Faiss (Disk-IVF) and Milvus (DiskANN). Our experiments show that Faiss offers 3.4× faster indexing and requires less storage, while Milvus provides 2.04× faster query throughput,allowing practitioners to choose between speed, storage efficiency, and query performance based on specific use cases.
    %\item We introduce a Type-1 de-duplication mechanism that indexes only a single representative embedding from each Type-1 clone class. By excluding redundant embeddings, this approach reduces index size and computational load, increasing the diversity of top-k results and improving retrieval precision for non-identical clones (Type-2, Type-3, and Type-4).

    %DAVID: I'm torn between putting global top k in as a contribution, and giving more focus to our other contributions where the quantitive evaluation can be seen more clearly.
    %\item We propose a Global Top-K strategy to consolidate and rank results across multiple queries, thereby prioritizing high-confidence clone pairs at the top of the results list. This approach maximizes recall while maintaining high precision at the top, allowing users to review the most relevant clone pairs first,improving both the efficiency and effectiveness of the clone detection process.
    
\end{itemize}

\noindent The rest of this paper is structured as follows: Section \ref{background} introduces clone detection problems and the challenges that need to be solved to search for clones in billion-line industrial code bases.
%reviews relevant background and prior work on code clone detection, including neural network-based methods, as well as nearest neighbor search techniques that enable scalability.
Section \ref{proposed-apporach} describes our solution to the problem of industrial scale clone detection, including specific challenges when dealing with very large code bases. Section \ref{experimental-setup} outlines the experimental setup, including details on the datasets, configuration parameters, and evaluation metrics used to assess both effectiveness and performance. Section \ref{results} presents an evaluation of our solution with respect to our two research questions: the ability of our solution to achieve similar accuracy to scalable in-memory solutions, and the ability of our solution to scale to industrial-size clone search problems with at least a billion lines of code. Section \ref{threats-to-validity} discusses potential threats to validity that could impact the generalizability of our findings. Finally, Section \ref{conclusion} draws conclusions and outlines potential directions for future research and improvements.

\setlist{topsep=0pt, leftmargin=*}

 \section{Background}
 \label{background}
 \subsection{Clone Comparison and Clone Search}
 There are two main sub-problems within automatic clone detection.
\textit{Clone comparison} involves directly comparing pairs of code fragments to determine if they are clones. Classical algorithmic methods such as token-based \cite{Kamiya2002654} and tree-based \cite{Jiang2007} comparisons have long been effective, but in recent years neural networks
have emerged that achieve very high levels of accuracy in clone
comparison. However, scaling clone comparison to large datasets remains a significant challenge. Since comparing $N$ code fragments requires $O(N^2)$ pairwise comparisons, this approach becomes impractical when working with thousands or millions of code fragments.

\textit{Clone search} on the other hand, is a more difficult problem where the goal is to find code clones among millions or even billions of lines of code. 
The input to the clone search consists of two parts. The \textit{corpus} consists of a large amount of existing code, such as billions of lines of open source code, or an existing corporate code repository. The \textit{query} is another (or the same) set of code. The code search problem is to find code clones that consist of pairs of code fragments, with one from each of the \textit{corpus} and \textit{query}. 

% Chochlov et al. \cite{Chochlov2022}, use the CodeBERT neural network to generate a fixed-length vector (or \textit{embedding}) for each code fragment. Then an efficient approximate nearest neighbour search is used to find code segments with similar embeddings. Thus, neural networks can be used to find similar code segments without pairwise comparison. 

\subsection{Scalable Clone Search: Beyond Pairwise Comparison}
%\textbf{This subsection could very briefly explain the SSCD approach [Done]}

To handle industrial-scale datasets, clone search systems face significant scalability challenges. Scalable Semantic Clone Detection (SSCD), introduced by Chochlov et al. \cite{Chochlov2022}, addresses these by moving beyond traditional pairwise comparisons. SSCD uses fine-tuned pre-trained models like CodeBERT and GraphCodeBERT to generate numerical embeddings that capture the semantic meaning of each code fragment. Instead of comparing every pair of fragments, SSCD performs a $k$-nearest neighbor ($k$-NN) search to identify the most similar fragments, significantly reducing the number of comparisons needed.
To further enhance scalability, SSCD employs an approximate $k$-NN ($k$-ANN) algorithm, which maintains high accuracy while improving search efficiency. For this, SSCD leverages in-memory indexing techniques, particularly the Hierarchical Navigable Small World (HNSW) graph, commonly used in approximate $k$-NN search \cite{Malkov2020, johnson2019billion}.

By using these advanced techniques, SSCD efficiently detects complex Type 3 and Type 4 clones (inexact or functionally similar clones), which are typically harder to identify. Its architecture allows it to scale effectively to much larger codebases, overcoming the bottleneck of pairwise comparisons and enabling clone detection in corpora containing hundreds of millions of lines of code.

\subsection{Industrial Scale Challenges in Clone Search}
%\textbf{This section could explain the challenges of industrial-size clone search, such as:
%\begin{itemize}
%    \item Choclov finds clones within a single code base, we want to have a separate CORPUS and QUERY [it can]
%    \item Choclov builds the index anew on each execution; we want it to be persistent [done]
%    \item We need a scalable search[done]
%\end{itemize}}

SSCD solves many important problems in making neural network clone detection scalable. It is particularly effective when the size of the corpus and query code are similar. However, it has some important limitations, especially when the corpus is very large and the query is small. In the particular real-world industrial case that we address, we have a large corpus of more than one billion lines of existing code. As software engineers within the company develop new software, we search for segments of new code that are clones for code segments within the existing corpus. In this scenario, the new code acts as a set of \textit{query} code that segments against a large \textit{corpus} of more than a billion lines of code. Thus, the \textit{corpus} is orders of magnitude larger than the \textit{query}.

SSCD uses the best in-memory algorithms for approximate $k$ nearest neighbor ($k$-ANN), such as Hierarchical Navigable Small World (HNSW). These methods build a search data structure in-memory to represent the \textit{corpus} and/or \textit{query} and allow fast, sub-quadratic nearest-neighbour search. Building the search data structures takes perhaps a majority of the execution time of these algorithms. If the \textit{query} and \textit{corpus} are different every time, then there is no alternative to building new search data structures on each execution of the tool. 
%% start of added text for camera ready version
A common case of the \textit{corpus} being different each time is clone detection where we search for clones within the code base of a project that is being actively modified. Detecting clones may be an important step in code refactoring, or in detecting clones of code that requires a bug fix while modifying the code base.
But in our industrial application, the \textit{corpus} is largely stable and it is only the \textit{query} that changes from one execution to another.

Thus, the SSCD approach of using a generic $k$-ANN algorithm that builds the index anew on each execution is wasteful, particularly for a billion-line \textit{corpus} where building the index can take several hours. Thus, to find a scalable solution for our industrial clone search problem, we need to make the embeddings and the search index for the \textit{corpus} persistent between executions of the clone search tool.

Another key limitation of relying on in-memory $k$-ANN algorithms is that the entire index must be resident in memory for efficient retrieval. This poses a scalability challenge when working with very large datasets. It is important to note that the memory requirements of the SSCD neural-network clone search method is much larger than the size of the source code alone. In the SSCD approach, a neural network creates a 768-dimensional vector embedding for each fragment of source code, which occupies around 3KB of storage. Although 3KB of storage for each embedding is small, we may need to store tens or hundreds of millions of embeddings. Further, data structures that allow nearest neighbours to be found efficiently among many millions of embeddings require further memory. Thus, a truly scalable clone search solution needs to be capable of efficiently indexing and searching on external storage.

% For example, embedding millions of code fragments into 768-dimensional vectors requires a significant amount of memory—approximately 460 GB of RAM for 150 million embeddings—far exceeding the capacity of most single-node machines. As a result, scaling SSCD to handle industrial-sized codebases necessitates either extremely high-end hardware or the use of distributed systems, both of which increase complexity and cost.

% Another limitation of SSCD is that it rebuilds the index with every execution, which is inefficient for large-scale applications. Rebuilding an index for millions or billions of code fragments is time-consuming and resource-intensive, making SSCD less suitable for frequent searches or real-time use cases. Persistent indexing would be necessary to improve its practicality and efficiency in such scenarios.

\begin{comment}
While both HNSW  and IVF indexes provides scalability for many use cases, they still require that the entire index fit into memory, which can become a bottleneck for very large datasets.
For example, a BERT-based embedding of 768 dimensions takes up 3KB of memory. Storing 150 million such embeddings requires about 460 GB of memory, far exceeding the capacity of a typical single-node machine. Thus, these memory-based approaches require multiple machines to scale, leading to increased complexity and cost.
\end{comment}
\section{The Proposed Approach: DB-SSCD}
\label{proposed-apporach}
For neural network clone search to be feasible at an industrial scale, with a \textit{corpus} of potentially billions of lines of code, we need a mechanism to perform nearest neighbour search on sets of code embeddings that exceed the size of memory. 
This section outlines the architecture of DB-SSCD that is  designed to enable scalable, disk-resident code clone detection for industrial-sized codebases. Our approach builds upon the workflow established by Chochlov et al. in SSCD, following a similar pipeline for embedding generation and clone search but differing in how we manage and index embeddings to handle significantly larger codebases. Note that although we use the term disk-based, our methods are equally applicable to solid state external storage. Indeed a fast solid-state drive will likely allow much faster clone search than a traditional hard disk.

\subsection{Architecture of DB-SSCD}

\begin{figure*}[tbhp]
\includegraphics[width=\textwidth]{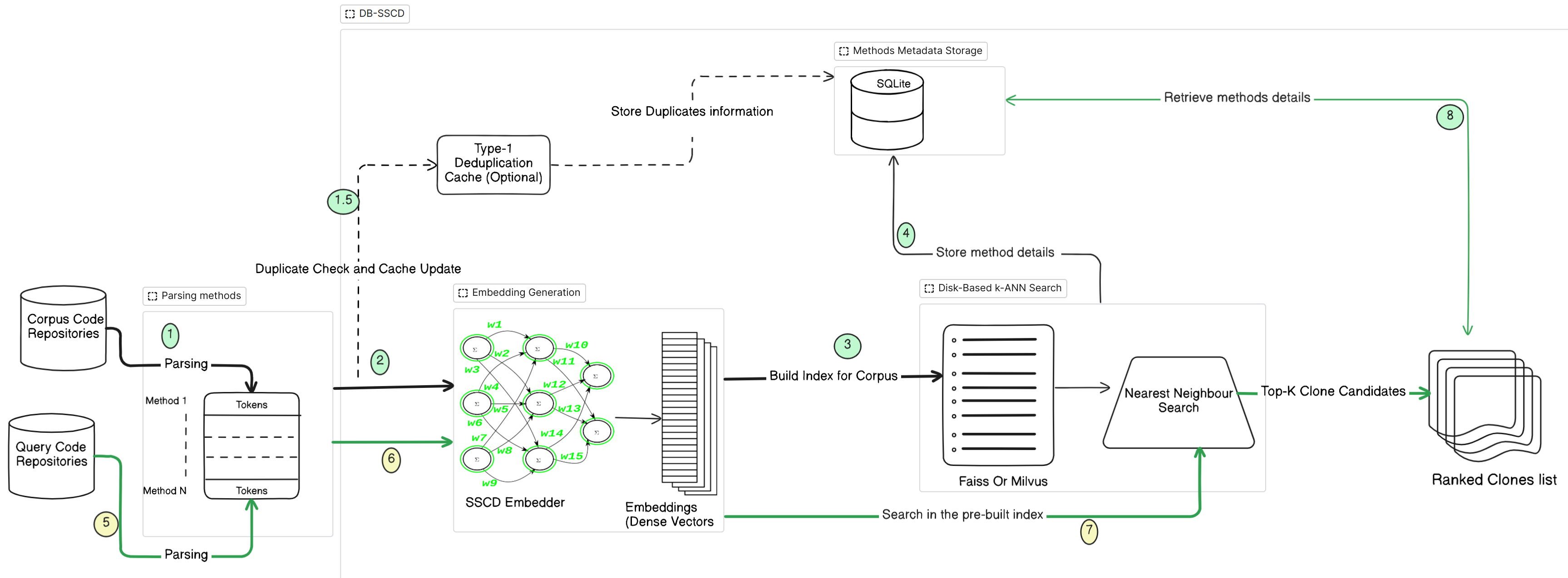}
\caption{DB-SSCD: Overall System Architecture for Clone Search}
\label{fig:sscd}
\end{figure*}

\noindent The DB-SSCD architecture includes two main processes: Corpus Indexing, a one-time setup, and Querying for Code Clones. The system’s overall architecture is shown in Fig. \ref{fig:sscd}.
\begin{enumerate}
    \item Corpus Indexing (One-time Process):
    \begin{itemize}
        \item \textbf{Parsing and Embedding Generation:} DB-SSCD begins by parsing the entire corpus of code repositories to extract individual methods (Step 1 in the diagram). Each method or code fragment is then processed by the SSCD Embedder—a CodeBERT-based model fine-tuned by Chochlov et al.\footnote{SSCD Embedder is available at https://huggingface.co/mchochlov/codebert-base-cd-ft}—to generate a fixed-length vector embedding. These embeddings represent each code fragment as a dense vector (Step 2).
        \item \textbf{Duplicate Check and Cache Update (Optional):} Before generating an embedding, DB-SSCD hashes each parsed method and checks this hash against a Type-1 Deduplication Cache (Step 1.5). If the hash is not in the cache, the method proceeds to embedding generation, and its hash is added to the cache. If it is already present, the method is recorded as a duplicate in the metadata database, and embedding generation is skipped. Details of this step are provided in Section \ref{subsec:type-1-deuplication}.
        \item \textbf{Building the Index:} The generated embeddings are indexed on disk using either Faiss or Milvus (Step 3). This indexing process is a one-time setup. Once created, the index can be stored on disk and reloaded as needed, avoiding repeated computational overhead and enabling efficient future queries.
        \item \textbf{Storing Method Information in Database:} When embeddings are added to the index (either Faiss or Milvus), they are assigned a unique identifier specific to that index. The metadata associated with each method—such as file path, start and end line—is stored in an SQLite database (Step 4) alongside this unique identifier. This setup allows DB-SSCD to efficiently retrieve and display the correct details for each method during querying, linking each embedding back to its exact source code location.
        
        %\item These embeddings are then stored on disk using large-scale indexing software. Our implementation offers two options for storing and indexing embeddings: Faiss and Milvus.
        
        %\item  We build a search index for the embeddings on disk using the indexing software. The index allows fast approximate nearest neighbor ($k$-ANN) search for similar embeddings. For a large \textit{corpus}, the size of the embeddings and index may far exceed available RAM. Therefore, the index may need to be built in multiple passes.
    \end{itemize}
%This embedding and indexing process only needs to be performed once and can be reloaded from disk for future queries, avoiding repeated computational overhead.
    
    \item Querying for Code Clones:  
    \begin{itemize}
        \item \textbf{Query Parsing and Embedding Generation:} In the query phase, DB-SSCD reads from query repositories specified by the user, similar to how it processes the corpus in the indexing phase (Step 5). The system parses the files within these query repositories, extracting methods from each file. Each extracted method is then passed through the SSCD Embedder to generate a dense vector embedding that represents the query code fragment (Step 6). This embedding is compatible with the embeddings generated for the corpus, allowing for a direct similarity comparison.
        \item \textbf{$k$-ANN Search on Disk-Based Index:} The query embedding is sent to the Disk-Based $k$-ANN Search module, which uses the Faiss or Milvus index to find the most similar embeddings in the corpus (Step 7). The $k$-ANN search returns the unique IDs of the top-$k$ nearest neighbors, representing the closest matching code fragments.
        \item \textbf{Query Results}: The unique IDs from the nearest neighbor search are used to look up the associated method metadata in the SQLite database (Step 8). This retrieval provides the file path and line number details for each matched code fragment, enabling users to locate the precise positions of potential clones. Finally, the matching code fragments are returned to the user as potential code clones.
        
        %\item This query vector is passed to the $k$-ANN search, which searches the on-disk index of the \textit{corpus} to find similar embeddings. 
        %\item The $k$-ANN search returns the vector IDs of the most similar code fragments. Using these vector IDs, DB-SSCD performs a metadata lookup in an SQL database to retrieve the corresponding file path and line numbers of the matched code.
        %\item Finally, the matching code fragments are returned to the user as potential code clones.
\end{itemize}
\end{enumerate}

\subsection{Disk-Based Indexing for Large-Scale Clone Search}
% $Could the description of indexing methods be described more in the context of our tool?$ addressed somewhat

To address the memory limitations of traditional in-memory indexing methods, DB-SSCD integrates two effective approaches to disk-based indexing, inspired by recent advancements in efficient search and indexing of large-scale high-dimensional vector data. Specifically, DB-SSCD employs a partition-based method using Faiss, similar in concept to SPANN \cite{Ranzato_NEURIPS2021}, and a graph-based approach through Milvus’s DiskANN \cite{Jayaram_NEURIPS2019}. Both methods leverage disk storage to manage datasets far exceeding available memory.

Faiss \cite{johnson2019billion}, a high-dimensional similarity search library developed by Facebook AI, offers several indexing options, including Inverted File (IVF) and Hierarchical Navigable Small World (HNSW) graphs. DB-SSCD leverages Faiss’s IVF index with an on-disk storage variant, designed to hold only cluster centroids in memory, while full embedding data is stored on disk in inverted lists.
The process begins with a so-called ``training'' step. In this step, a random sample of embeddings are clustered using a $k$-means algorithm to identify centroids representing different partitions of the multi-dimensional embedding space. These centroids are stored in memory, while the full embeddings are stored in on-disk inverted lists associated with each cluster. To avoid memory overflow, embeddings are indexed in batches. Each batch is processed, and the indexed segments are saved to disk as independent blocks. After all batches have been processed, the segments are merged into a unified index, consolidating the inverted lists for more efficient retrieval.

During search, the centroid index is kept in main memory, allowing DB-SSCD to quickly narrow down the relevant clusters for a query. DB-SSCD specifies the number of nearby clusters to search for a nearest-neighbour using the $nprobe$ parameter, balancing speed and search precision. By selectively loading only the necessary inverted lists from disk, Faiss enables DB-SSCD to conduct high-speed, large-scale searches without loading all embeddings into memory, making it possible to search in vast codebases efficiently.

In addition to the Faiss-based partitioning approach, DB-SSCD integrates a graph-based indexing method through Milvus's DiskANN. Milvus \cite{Wang_Milvus_2021}, an open-source vector database system, supports the DiskANN indexing structure, which efficiently stores embeddings as nodes within a sparse proximity graph on disk. DiskANN organizes vectors into a navigable graph structure where edges represent proximity in the vector space. The graph structure and full vector data reside on disk, while compressed versions of vectors are maintained in memory using Product Quantization \cite{jegou2010product}. During search, DiskANN first uses these compressed vectors for approximate distance calculations while traversing the graph, only loading full vectors from disk when higher precision is needed.

Our decision to choose Milvus was primarily influenced by its support for DiskANN, a critical feature for our requirements. Additionally, we sought to adopt a vector database due to their increasing prominence in handling high-dimensional vector data. After evaluating solutions based on Faiss, we aimed to integrate a solution that exemplifies the vector database ecosystem, and Milvus was the only system that supported DiskANN at the time, making it the most suitable choice for our needs.

By utilizing both systems, we can explore and evaluate two
distinct types of on-disk indexing and searching for clone search: one based
on the Inverted File (IVF) system, and the other rooted in
graph-based methodologies.

\subsection{Processing Results}
%How do we decide which results to show the user, and in what order? [Done]

Research studies of clone detection algorithms often seek to find all clones within a single code base. In our industrial setting, we have a slightly different clone search problem. We have a very large corpus of more than a billion lines of existing code. As we develop new code, we want to detect whether the new \textit{query} code fragments are possible clones of any existing fragments in the corpus. To do this, we use a $k$-ANN ($k$-Approximate Nearest Neighbor) search to identify the $k$ most similar corpus code fragments for each query fragment.

For each query fragment, the $k$-ANN search yields a list of the top \( k \) most similar results based on cosine similarity scores. When dealing with a batch of queries, denoted by \( Q = \{q_1, q_2, \dots, q_m\} \), the $k$-ANN search returns a list of top-$k$ results for each individual query:
\[
R_i = \{ q_i r_1, q_i r_2, \dots, q_i r_k \}
\]
where each result \( q_i r_j \) for query \( q_i \) is ranked by its similarity score \( s(q_i, r_j) \), such that \( s(q_i, r_1) \geq s(q_i, r_2) \geq \dots \geq s(q_i, r_k) \). This Per-Query Top-K approach provides the user with a separate ranked list of results for each query, enabling them to view the closest matches for each individual query.

In an industrial setting where multiple code repositories are processed together (i.e., multiple queries) to find clones, we found specific challenges with this approach. Similarity scores across different queries vary widely, meaning that the top-$k$ results of some queries may contain highly similar results, while others may include lower-quality matches. In such cases, the user may need to review results with widely varying similarity scores across queries, which increases the likelihood of encountering false positives for queries with inherently lower similarity scores. Users have to filter out less relevant results from each query’s list, resulting in unnecessary overhead and potentially missed clones. In clone detection, the goal is to identify the most relevant matches across all queries, not just within each query.

\textbf{Global Top-K}: To address these issues, we propose the Global Top-K strategy. This approach aims to enhance the relevance and consistency of results across a batch of queries by consolidating and re-ranking results globally based on their similarity scores.

The Global Top-K strategy involves three main steps:

\paragraph{Retrieve Top-K Results per Query}
For each query \( q_i \), obtain a list of the top \( k \) results based on similarity scores:
\begin{equation}
\begin{aligned}
R_i = \{ q_i r_1, q_i r_2, \dots, q_i r_k \}
\end{aligned}
\end{equation}
This results in a batch of results across all queries where \( m \) is the batch size:

\begin{equation}
\begin{aligned}
R_{\text{batch}} &= \{ R_1, R_2, \dots, R_m \} \\
&= \{ \{ q_1 r_1, q_1 r_2, \dots, q_1 r_k \}, \\
&\quad \dots, \{ q_m r_1, q_m r_2, \dots, q_m r_k \} \}
\end{aligned}
\end{equation}

This initial retrieval phase gives us \( m \times k \) results.

\paragraph{Aggregate and Sort All Results by Similarity Score}
Combine all individual query results into a single list:
\[
R_{\text{all}} = R_1 \cup R_2 \cup \dots \cup R_m
\]
Sort \( R_{\text{all}} \) by relevance scores so that the highest relevance results appear first, regardless of the originating query. Let this sorted list be:
\begin{equation}
\begin{aligned}
R_{\text{sorted}} &= \{ r_{(1)}, r_{(2)}, \dots, r_{(mk)} \} \\
&\text{where} \quad s(r_{(1)}) \geq s(r_{(2)}) \geq \dots \geq s(r_{(mk)})
\end{aligned}
\end{equation}

\begin{figure}[tbhp!]
\includegraphics[width=\columnwidth]{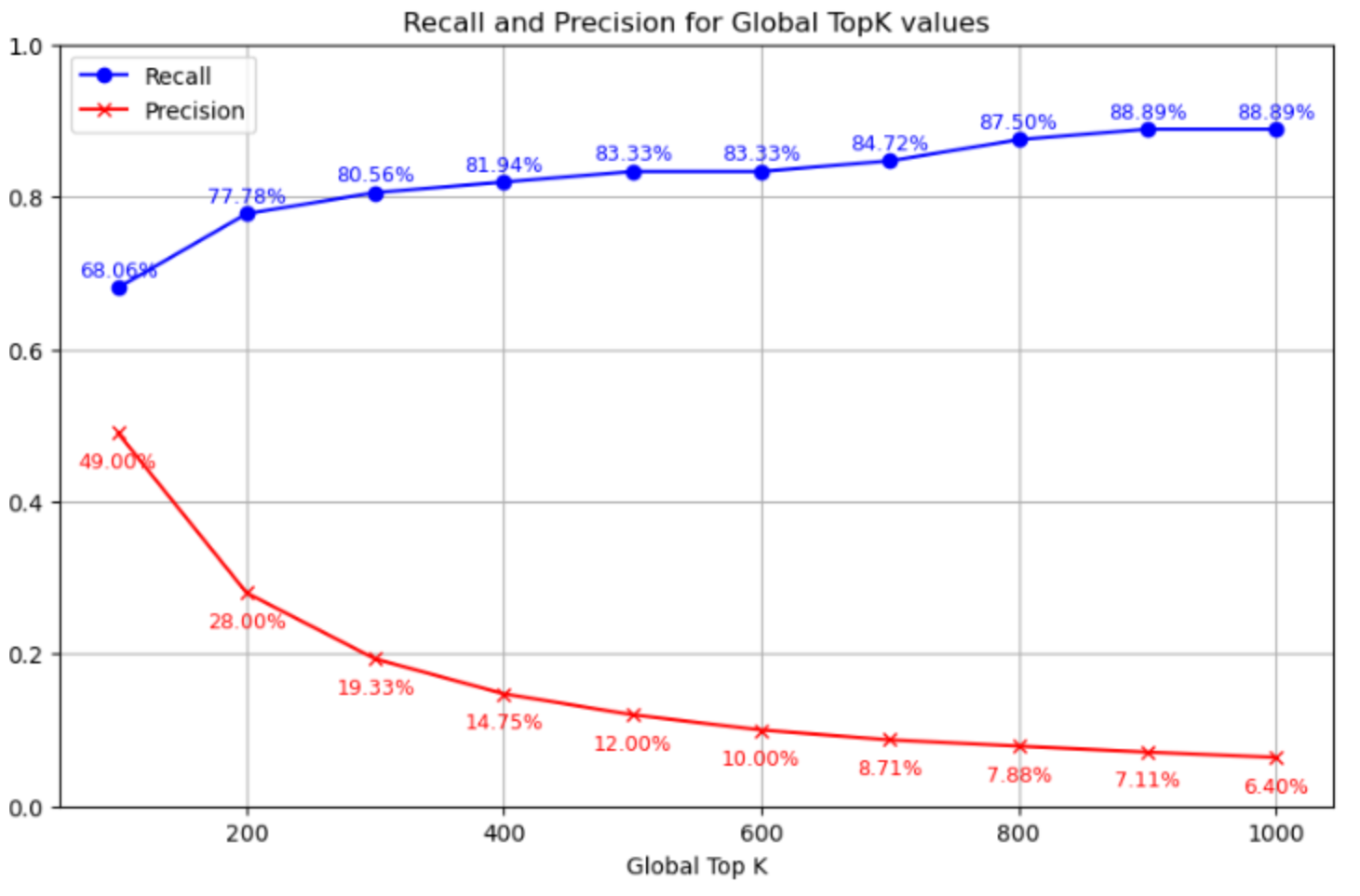}
\caption{Global Top-K Strategy: Recall and Precision Trade-off for Global Top-K on the Company-C Dataset.}
\label{fig:gbl-topk-illustration}
\end{figure}

\paragraph{Select the Global Top-K Results}
From \( R_{\text{sorted}} \), select the top \( k_{\text{global}} \) results:
\begin{equation}
\begin{aligned}
R_{\text{global-top-k}} = \{ r_{(1)}, r_{(2)}, \dots, r_{(k_{\text{global}})} \}
\end{aligned}
\end{equation}
This final list, \( R_{\text{global-top-k}} \), represents the Global Top-K results across all queries and is presented to the user. This Global Top-K approach ensures that only the highest relevance results across all queries are presented. By focusing on the global ranking, users are more likely to see true clones or highly relevant matches at the top, minimizing the time they spend filtering out false positives.
% Per-Query Top-K (Problem):Users may see a mix of high and lower-similarity results, possibly with more false positives due to inconsistent relevance across queries. (m x k results) 
% Global Top-K (Advantage):Users see the most relevant results across all queries, with a higher likelihood of true positives (reduced false positives). (k-glboal results)
%Fig. \ref{fig:gbl-topk-illustration} demonstrates the recall and precision trade-off for the Global Top-K strategy on a C language clone detection dataset\footnote{https://github.com/SFI-Lero/SSCD}.
Fig. \ref{fig:gbl-topk-illustration} illustrates the recall and precision trade-off for the Global Top-K strategy, using a C language clone detection dataset. Provided by our industrial collaborators and referred to in SSCD \cite{Chochlov2022} as the Company-C dataset, it is publicly available on GitHub\footnote{https://github.com/SFI-Lero/SSCD}.The dataset contains approximately 80,190 lines of code spread across 1,714 methods within 160 files and includes 80 manually identified clone pairs representing various clone types.

The Fig. \ref{fig:gbl-topk-illustration} plot highlights that as the Global Top-K value increases, recall improves steadily, reaching nearly 89\%, indicating that the strategy is successfully retrieving a large portion of all true clone pairs present in the dataset. At lower Global Top-K values, precision is notably high, showing that the strategy prioritizes the results that the model is most confident about. This high initial precision suggests that true clone pairs are effectively ``bubbled up'' to the top of the list, minimizing false positives in the early results. This precision-recall balance allows users to either prioritize a concise list of highly relevant clones or expand the scope to include additional candidates, offering flexibility depending on the needs of the clone detection task.
%% Change for final camera ready version
Whether precision or recall is more important depends on the clone detection task. For example, if the task is to refactor instances or identical or similar code, then precision is likely to be the most important factor. It is not a problem to miss some potentional clones, and indeed the less similar are the clones, the less benefit is likely to arise from refactoring them. In contrast, in a legal case about the authorship of code, we might want to prioritise recall. In this case we might  be willing to manually process several false-clone candidates for each true clone that we find.

 \subsection{Type-1 Deduplication}
 \label{subsec:type-1-deuplication}
A key parameter of $k$-nearest neighbour algorithms is the number of neighbours, $k$, that should be found for each code fragment embedding in the query. The execution time of the $k$-NN algorithm increases with $k$, so finding a large number of nearest neighbours for each embedding in the query can have a high computational cost.

Type 1 clones are pairs of code fragments that are identical in all respects except for differences in indentation or white space. It is not unusual for identical or near-identical code fragments to appear multiple times in a large program. In an industrial-size code base containing many programs, we may see very large numbers of identical code fragments. For example, in the BigCloneBench (BCB) clone detection dataset \cite{Svajlenko2017a}, we observe some code fragments that appear more than a thousand times. These very large groups of identical code fragments cause problems for $k$-nearest neighbour search when the top-$k$ nearest items are all members of one of these large groups of Type-1 clones.
 
\begin{figure}[tbhp]
\includegraphics[width=\columnwidth]{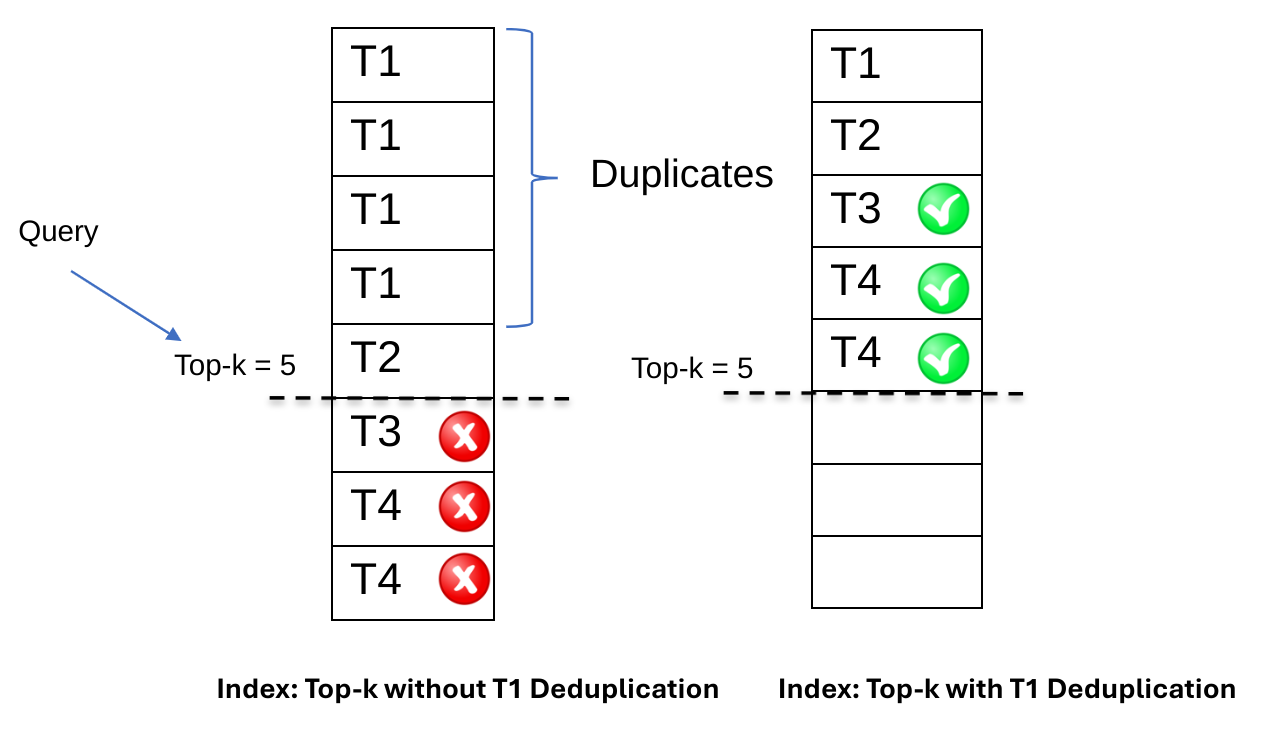}
\caption{Illustration of the Impact of Type-1 Clone Deduplication on Top-k Candidate Diversity}
\label{fig:type-1-dedup}
\end{figure}

 Typically, nearest-neighbor search-based systems return a fixed number of candidates for each query, determined by the parameter top-$k$. When exact code duplicates (Type-1 clones) exist across a codebase, they can significantly affect search quality by dominating the limited top-$k$ result slots. For example, in a system that returns the top 5 most similar code snippets, multiple identical copies of the same code (forming a clone class) can occupy several, if not all, of these slots. This redundancy not only increases computational overhead by requiring the system to index and search through multiple embeddings of the same code, but it also diminishes the user's ability to discover diverse and potentially more relevant clone sets.  This issue is illustrated in Fig. \ref{fig:type-1-dedup}  (left), where the top-$k$ results are dominated by identical clones.
To address this issue, DB-SSCD implements an efficient deduplication strategy for handling Type-1 clones. When the system encounters a new code fragment, it first checks whether an identical embedding has already been indexed. If a match is found, the system bypasses indexing the duplicate and instead maintains only a reference to the original embedding in the database. This ensures that while only one instance of the Type-1 clone is indexed, all duplicates remain accessible to the user at search time. In the example shown in Fig. \ref{fig:type-1-dedup}  (right), along side of Type-1 the unique clones such as T2, T3, and T4 can be returned in the top-$k$ results now. This approach enhances search efficiency and improves the overall user experience by ensuring that Type-1 clone duplicates remain accessible, while optimizing the top-$k$ results to include a broader variety of clones.

\section{Experimental Setup}
\label{experimental-setup}
In this section, we describe the experimental setup used to evaluate the performance, scalability, and accuracy of DB-SSCD. Our experiments are designed to assess the system’s ability to scale to large datasets that exceed the available memory, while maintaining high recall and efficiency in code clone detection.
The specifications of the machine used for the evaluation are summarized in Table \ref{tbl:machines_spec}. The available RAM is 32 GB, but all experiments were conducted by deliberately limiting the RAM to 16 GB to simulate a real-world environment with limited resources, demonstrating the ability of both Faiss and Milvus to handle datasets that far exceed available memory. Note that the ``disk'' that we use in our experiments is not a hard disk, but a fast 2TB NVMe solid-state drive.

\begin{table}[t]
\centering
\caption{Machine Specifications}
%\resizebox{\columnwidth}{!}{
\small
\begin{tabular}{|c|p{1.8cm}|p{0.7cm}|p{0.9cm}|p{1.5cm}|}
\hline
\textbf{Machine} & \textbf{Processor} & \textbf{RAM} & \textbf{Storage} & \textbf{GPU} \\
\hline
M1 & AMD Ryzen 5 3600X 6-Core & 32 GB & 2 TB NVMe SSD & 1x 12 GB GeForce RTX 2080  \\
\hline
\end{tabular}
%}
\label{tbl:machines_spec}
\end{table}

\textbf{Datasets.} Two common benchmarks for code clone detection are the BigCloneBench\cite{Svajlenko2017a} and POJ-104 \cite{Mou2016} datasets. However, there is currently no code clone detection dataset large enough to test industrial corpus sizes. The embedding size of Full-BCB dataset is about 15 GB, which can be processed easily on a typical system with 16 GB of RAM. %Therefore, we instead use the The Stack dataset \cite{Kocetkov2022TheStack} from the BigCode project (bigcode-project.org). This dataset includes 6.4 TB of   permissively licensed source code in 358 different programming languages.
To evaluate the scalability of DB-SSCD on industrial-sized codebases, we used the C programming language subset of the BigCode Stack dataset \cite{Kocetkov2022TheStack}, which contains 1.16 billion lines of code. The Stack includes permissively licensed source code from various open-source repositories and consists of millions of files and functions. Although this dataset does not contain clone labels, it allows us to measure the system’s indexing and query performance on a massive scale. However, without labels, we cannot determine the precision or recall of clone detection.

\begin{table}[tbp]
\centering
\caption{Summary of BigCode ``Stack'' C Source Code statistics, Parsing Time, and Embedding Generation }
\small
\begin{tabular}{|l|r|}
\hline
\textbf{C-Dataset} & \textbf{Values} \\
\hline
\# Original Lines of Code & 1,160,433,488\\
\# Original Files & 8,625,559 \\
\# Cleaned Files & 2,735,164 \\
\# C Functions Parsed & 18,173,452 \\
Parsing Time & 64 mins \\
\# Code Embeddings & 18,173,452 \\
Embedding Generation Time & 25h 17m \\
Size On Disk (embeddings)  & 52 GB \\
\hline
\end{tabular}

\label{tbl:dataset_summary}
\end{table}

We created a 52 GB corpus from the C code in The Stack, large enough to challenge systems that must scale beyond memory limits and test their ability to handle indexing and querying of datasets significantly larger than available memory.

A summary of the dataset is shown in Table \ref{tbl:dataset_summary}. There are 8,625,559 files in the dataset and after cleaning the dataset i.e removing non parsable files, binary files and files that do not have .h and .c as an extension. There were 2,735,164 remaining files with 18,173,452 C functions parsed. Parsing took 64 minutes to complete. Running the CodeBERT neural network on machine M1 to compute embeddings of these 18 million functions took 25 hours and 17 minutes. The  size of the embeddings is 52 GB, which is more than thrice the allocated RAM for the experiments. The \textit{corpus} code is normally a relatively fixed database of code that we prepare once and query many times.

\textbf{Indexing and Searching Parameters}
\label{index_search_parameters}
We set up a collection within the \textbf{Milvus} database, using DiskANN as the index, and the  L2 (Euclidean) distance metric. The two parameters that control the behavior of the graph algorithm in DiskANN are $R$ and $L$. $R$ (default 56) specifies the maximum number of neighbors (out-degree) each node can have in the graph. It ensures that the graph remains sparse by limiting the number of connections per node, which reduces the amount of data that needs to be read from disk for each node. $L$ (default 100) controls the size of the candidate list maintained during the greedy search. A larger candidate list size allows more nodes to be considered, potentially improving search accuracy at the cost of increased computational effort.

We construct a \textbf{Faiss} disk-based IVF index without compression, specifically an IVF\_FLAT index. The construction of the IVF index requires an additional clustering step to find centroids of Voronoi cells within the dataset. Faiss refers to this clustering step as ``training'' the index.
To determine the suitable number of clusters for a given dataset, we used a heuristic value suggested in the Faiss documentation: $\text{num\_of\_clusters} = 8 \times \sqrt{N}$. For The Stack Dataset \( N = 18,173,452 \), and based on this heuristic,  34,104 clusters were created. The sample size for training the $K$-means clustering is determined to ensure a representative subset of the entire dataset. Faiss recommends $\text{sample size} = 39 \times \text{num\_of\_clusters},$ where 39 is the minimum number recommended per cluster. Consequently, we used a random sample of 1,330,056 vectors from the dataset for training.
The parameter $nprobe$ is used to specify how many different \textit{clusters} are searched for the nearest neighbours in each query. We dynamically select $nprobe$ based on the number of clusters in the index, using the heuristic $ \textit{nprobe} = \sqrt{\textit{num\_of\_clusters}}.$ Since the number of clusters in the Stack dataset was set to 34,104, the $nprobe$ is calculated to be 184. This means that when searching for nearest neighbors, we search within the 184 closest clusters.

\subsection{Evaluation Metrics}
We evaluated DB-SSCD based on the following metrics.
\begin{itemize}
    \item  \textbf{Indexing Time:} The time taken to construct the disk-based index from the embeddings.
    \item \textbf{Queries per Second (QPS):} The number of queries that can be processed per second, which measures the system's throughput.
    \item \textbf{Recall:} For labeled datasets such as BigCloneBench, we measured the recall of clone detection to evaluate how accurately DB-SSCD identifies clones compared to in-memory methods.
    \item \textbf{Disk Usage:} The amount of disk space required to store the embeddings and the index.
    \item \textbf{Memory Efficiency:} The system's ability to operate effectively within imposed memory constraints, demonstrated by successfully indexing and querying a dataset significantly larger than the available memory limit.
\end{itemize}

\subsection{Experiment Design}
\subsubsection{Indexing}
The first set of experiments involved indexing the BigCloneBench and The Stack datasets using both Faiss and Milvus. We measured the total time taken to build the index, as well as the time per embedding. 
 \subsubsection{Search Performance}
The second set of experiments focused on evaluating the query performance of DB-SSCD. We generated 10,000 random query embeddings from each dataset and measured the time taken to retrieve the top 10 nearest neighbors. For each system, we calculated the average query time per embedding and the queries per second (QPS) to assess throughput.

\subsubsection{ Recall for Clone Detection}
To ensure that DB-SSCD maintains high clone detection accuracy, we measured the recall on the BigCloneBench dataset. We compared the recall achieved by DB-SSCD (using disk-based indexing) with the results of Chochlov et al., who evaluated clone detection using an in-memory approach. The goal was to verify that disk-based search does not sacrifice recall for scalability.

\section{Results}
\label{results}
In this section, we present the results of the experiments designed to evaluate the scalability, performance, and accuracy of DB-SSCD using both Faiss and Milvus as the underlying disk-based nearest neighbor search systems. 
\subsection{RQ1: Can our solution achieve clone search accuracy similar to existing scalable clone detection tools? }
%\subsection{Recall on BigCloneBench}
% $Could we put this at the start of the results?$[move here]
To answer this question, DB-SSCD was evaluated on the BigCloneBench dataset, focusing on recall to assess accuracy in identifying similar code fragments. Table \ref{tbl:recall_bcb_full} shows a comparison of recall achieved by DB-SSCD with Faiss and Milvus indexing, alongside the recall reported by Chochlov et al. \cite{Chochlov2022}, which used an in-memory search system.

\begin{table}[th]
\renewcommand{\arraystretch}{1.5}
\setlength{\tabcolsep}{5pt} % Default value: 6pt
\caption{Recall Comparison of DB-SSCD with SSCD and Other Scalable Clone Detection Tools on  BCB-FULL dataset}
\label{tbl:recall_bcb_full}
\centering
\begin{tabular}{l|rrrrrc}
\hline
\multirow{2}{*}{Indexing Method} & \multicolumn{6}{c}{Recall (\%)} \\ \cline{2-7}
                     & T1  & T2  & VST3 & ST3 & MT3 & WT3/T4 \\ \hline
DB-SSCD (Faiss)           & 100  & 97  & 96   & 81  & 28   & 1      \\ \hline
DB-SSCD (Milvus)     & 99  & 96  & 96   & 80  & 26   & 1      \\ \hline
\hline
SSCD (In-Memory) \cite{Chochlov2022}             & 100 & 97  & 96   & 80  & 27  & 1      \\ \hline
SourcererCC   \cite{Chochlov2022}           & 85  & 94  & 52   & 52  & 2   & 0      \\ \hline
\begin{tabular}[c]{@{}l@{}}SAGA \cite{Chochlov2022} \end{tabular}                  & 100 & 100 & 95   & 60  & 10  & 0      \\ \hline
\end{tabular}
% (Milvus tk=100, distance=L2, Th<=25, Other Tk=100, distance=cosine, th>=0.95

\end{table}
For consistency in evaluation, DB-SSCD was configured using the same core parameters as SSCD, such as input length, top-$k$ nearest neighbors, and similarity threshold, to maintain comparability with previous results. Specifically, embeddings were generated with a 128-token input length, the cosine similarity threshold was set to 0.95, and top-$k$ was set to 100. This relatively high top-$k$ value was chosen to account for BigCloneBench's unique structure, where some code fragments belong to large clone classes containing many similar clones. In the SSCD evaluations, recall increased significantly when top-$k$ was raised from 10 to 100, indicating that a smaller top-$k$ would fail to capture all relevant clones within these extensive classes. The primary difference in DB-SSCD’s configuration lies in the disk-based indexing methods, Faiss (IVF) and Milvus (DiskANN), which introduce additional parameters, such as the number of clusters in IVF for Faiss and search graph configurations for Milvus (details of which are discussed in the Section \ref{index_search_parameters}).

%% Start change for camera-ready version
As shown in Table \ref{tbl:recall_bcb_full},
DB-SSCD using Faiss achieves recall levels that are comparable to the in-memory approach  across all clone types. For Type 1 and Type 2 clones, Faiss matches the recall of the SSCD approach. 
%% Start change for camera-ready version
%For Type 3 clones, Faiss even slightly improves the recall (by 1\%).
For strong Type 3 (ST3) and medium Type 3 (MT3), Faiss achieves slightly higher recall (81\% versus 80\% and 28\% versus 27\%, respectively). It is important to note that Faiss, Milvus and in-memory SSCD all find similar embeddings using approximate nearest-neighbour algorithms. Slight differences in the approximate algorithms can cause each tool to return slightly different sets of nearest neighbours. Further, in the BigCloneBench dataset, which contains hundreds of millions of lines of code, distances between different pairs of code fragments can sometimes be almost identical, to the point that minor differences in floating-point rounding can result in changes in the similarity-ranking of clone pairs. These two effects cause minor variations in the ranking of clone pairs from different tools. These minor variations account for the slightly higher recall achieved by DB-SSCD with Faiss.
The recall achieved by Milvus (DiskANN) is similarly high, closely matching the results from both Faiss and the in-memory approach. 
%% Start change for camera-ready version
%The recall drops slightly for Type 1 and Type 2 clones but remains highly competitive for Type 3 and Type 4 clones.
For very strong Type 3 (VST3), strong Type 3 (ST3) and weak Type 3/Type 4 (WT3/T4) the recall is that same as the original in-memory SSCD.  But for Type 1 (T1), Type 2 (T2), and medium Type 3 (MT3) Milvus has very slightly lower recall. This is again the result of slight variations in the similarity ranking caused by different approximate nearest neighbour algorithms.
%%

%For completeness, we report the results of two other scalable clone detection tools, SourcerCC \cite{Sajnani2016} and SAGA \cite{Li2020}, as presented in the work by Chochlov et al. \cite{Chochlov2022}
For context, we also include the recall results of two other scalable clone detection tools, SourcerCC \cite{Sajnani2016} and SAGA \cite{Li2020}, as compared to SSCD by Chochlov et al. \cite{Chochlov2022}. While both SourcerCC and SAGA demonstrate strong recall for exact and very similar clones, they are less effective for complex, functionally similar clone types (MT3 and WT3/T4).
Overall, DB-SSCD retains the high recall levels of in-memory SSCD while providing the scalability needed for disk-based indexing. 
\subsection{RQ2 Can our solution scale to search for clones in 
industrial-sized code bases with at least a billion lines of 
code?}
To answer this question, we assess DB-SSCD’s scalability by including an additional large dataset, The `Stack'', alongside BigCloneBench. This allows us to evaluate DB-SSCD’s ability to operate on codebases that reach billions of lines of code, far beyond the scale handled by traditional in-memory methods. 
\subsubsection{scalability}

Table~\ref{tbl:time_bcb_full} (adapted from Chochlov et al.) compares the total execution times of various scalable clone detection tools on the BCB-FULL dataset. We reproduce this table to contextualize the performance of DB-SSCD alongside established tools, particularly SSCD, from which DB-SSCD is derived.

DB-SSCD achieves execution times of 20h 25m with Faiss and 26h 49m with Milvus, which, although slower than SSCD’s in-memory approach, represents a practical trade-off for the ability to handle datasets beyond RAM constraints. The slower performance of DB-SSCD is largely due to disk access times, an inherent limitation of disk-based indexing; however, it remains competitive relative to other scalable clone detection tools. When considering  other tools that support large datasets, DB-SSCD is slower than the single-GPU SAGA but significantly faster than SourcererCC, CCFinder, and Oreo. 

The difference in execution times between DB-SSCD (Faiss) and DB-SSCD (Milvus) reflects a balance between indexing speed and query throughput, with Faiss offering faster indexing while Milvus supports higher query performance. 
 %Although DB-SSCD incurs additional processing time due to disk access, it successfully scales clone detection to codebases exceeding RAM limits while maintaining competitive execution times.
\begin{table}[!th]
\renewcommand{\arraystretch}{1.2}
\caption{Total execution time comparison for scalable Clone Detection Tools on the BCB-FULL dataset}
\label{tbl:time_bcb_full}
\centering
\resizebox{\columnwidth}{!}{
\begin{tabular}{lr}
\hline
CDT                                        & Total time  \\ \hline
\multicolumn{1}{l|}{DB-SSCD (Faiss, (CPU)), 1 GPU inference)}     & 20h 25m \\
\multicolumn{1}{l|}{DB-SSCD (Milvus, (CPU)), 1 GPU inference)}      & 26h 49m \\ \hline \hline
\multicolumn{1}{l|}{SSCD (exact, 1 GPU)}            & 11h 01m \\
\multicolumn{1}{l|}{SSCD (exact, 5 GPUs)}            & 2h 53m      \\
\multicolumn{1}{l|}{SourcererCC (CPU)}                    & 2d 20h 25m  \\
\multicolumn{1}{l|}{CCFinder (CPU)}                       & 2d 00h 52m  \\
\multicolumn{1}{l|}{CloneWorks (CPU)}                  & 15h 12m     \\
\multicolumn{1}{l|}{Oreo (CPU)}                            & 17d 16h 11m \\
\multicolumn{1}{l|}{SAGA (1 GPU)}                            & 2h 40m      \\ \hline
\end{tabular}

}
\end{table}
 \subsubsection{Indexing Performance}
Table \ref{tbl:faiss_milvus_combined} shows the indexing time for both Faiss (Disk-IVF) and Milvus (DiskANN) across two datasets: BigCloneBench and The Stack.
\begin{table*}[!htbp]
\renewcommand{\arraystretch}{1.5}
\centering
\small
\caption{Comparison of Disk-based indexing and searching built on top of Milvus and Faiss}
\label{tbl:faiss_milvus_combined}
\begin{tabular}{|c|c|c|c|c|c|c|}
\hline
\textbf{Dataset} & \multirow{2}{*}{\textbf{System}} & \multicolumn{1}{c|}{\textbf{Indexing}} & \multicolumn{4}{c|}{\textbf{Search}}  \\
                &               & total time       & number of queries & $topK$ & total time & QPS \\
\hline
Full BCB & Milvus & 137 mins & 10,000 & 100 & 63 seconds & 158.73/s \\
\hline
Full BCB & Faiss  & 32 mins  & 10,000 & 100 & 11.88 seconds & 841.75/s  \\
\hline
The Stack & Milvus & 646 mins & 10,000 & 10 & 135 seconds & 74.07/s   \\
\hline
The Stack & Faiss  & 189 mins & 10,000 & 10 & 275.9 seconds & 36.25/s  \\
\hline
\end{tabular}
\end{table*}

For both datasets, Faiss demonstrates faster indexing than Milvus. It is 4.28× faster than Milvus for indexing the BCB dataset and 3.42× faster than Milvus on The Stack dataset. Despite The Stack being a much larger dataset, Faiss continues to demonstrate superior performance in terms of indexing speed, which makes Faiss more suitable when fast indexing is critical. While Milvus is slower at indexing compared to Faiss, its DiskANN-based graph structure allows for more robust search operations, as discussed in the next section. The trade-off is the longer indexing time due to the construction of the graph-based structure.

\subsubsection{Query Performance}
We  performed a search operation using a set of 10,000 code fragment \textit{queries}. For each \textit{query} embedding we find the 10  nearest neighbours in the \textit{corpus} in The Stack dataset and 100 nearest neighbours in the Full-BCB Dataset. Table \ref{tbl:faiss_milvus_combined} also shows the query performance based on number of queries, topK (nearest neighbors), total search time, and queries per second (QPS).

Faiss outperforms Milvus in terms of query speed, particularly on the BigCloneBench dataset, with 5.3× higher query throughput (QPS). The Full-BCB dataset is about 15 GB, which is close to the system's 16 GB of RAM. This allows most of the data to remain in memory and require less movement of data to and from  disk. The structure of the IVF index enables efficient data reuse. Once the posting list for a given centroid is loaded into memory for a query, it does not need to be reloaded for subsequent queries, as long as there is enough memory and no eviction is necessary. This minimizes disk I/O, reducing query latency and improving overall performance. On The Stack dataset, Milvus outperforms Faiss, achieving 2.04× higher query throughput (QPS). 
Our evaluation shows that both Faiss and Milvus enable search on datasets exceeding available RAM, which in-memory systems cannot handle. Although their QPS (Faiss: 36.25, Milvus: 74) is lower than in-memory searches (which often achieves QPS in the thousands), their disk-based capabilities make them essential for handling large datasets where memory is insufficient.

\subsubsection{Disk Usage }
As shown in Table \ref{tbl:disk_usage}, there is a considerable difference between the disk space usage of both systems. Recall that the total size of the raw data, is approximately 52 GB. Faiss stores the data across two primary files: the index file and the file containing \textit{inverted lists}. The index file is compact, occupying 100 MB, as it mainly contains the core index structure and the collections of centroids, but not the actual vector data. The file containing the inverted lists, which includes the actual embeddings, is 53 GB in size. Thus, the additional disk space consumed by Faiss, beyond the raw data size, is about 1.02 GB.

Milvus, on the other hand, uses three key components for its data management. The Milvus Core, which serves a similar purpose to Faiss, takes up 134 GB of disk space. The MinIO component, which handles the storage of raw vector data and insertion logs, consumes 116 GB of disk space. Lastly, etcd, which is responsible for managing the metadata for Milvus's internal components like proxies and index nodes, uses 8.0 KB of disk space. In total, Milvus consumes 250 GB of disk space. Thus, the additional disk space consumed by Milvus, beyond the raw data size, is 198 GB.
\begin{table}[!h]
\centering
\caption{Disk Usage for Faiss and Milvus for The Stack dataset}
\resizebox{\columnwidth}{!}{
\begin{tabular}{|c|c|c|c|}
\hline
 \multicolumn{2}{|c|}{Faiss} & \multicolumn{2}{c|}{Milvus} \\
\hline
 Component & Size & Component & Size \\
\hline
Trained Index File & 100 MB & Milvus Core & 134 GB \\
\hline
Inverted Lists File & 53 GB & MinIO & 116 GB \\
\hline
Populated index & 101 MB & etcd & 8.0 KB \\
\hline
Total & ~53.02 GB & & 250 GB \\
\hline
\end{tabular}
}
\label{tbl:disk_usage}
\end{table}

%% Adding this text back in that was previously cut, presumably to address space problems
In summary, DB-SSCD demonstrates scalability to handle large-scale codebases that exceed memory limits, providing disk-based alternatives for indexing and searching through Faiss and Milvus. Although slower than in-memory approaches, DB-SSCD’s ability to operate efficiently on datasets surpassing a billion lines of code highlights its practical value for industrial applications where data size presents a barrier to conventional clone detection tools.

\section{Threats to Validity}
\label{threats-to-validity}
This work reports indexing and search times for two different types of indexing methods and software systems. The measured performance values can be influenced by various factors, including hardware configurations (e.g., CPU speed, disk speed, and RAM) and the specific versions of Faiss and Milvus used. Additionally, the choice of parameters for each indexing technique can significantly impact the results. For instance, in the IVF index, the number of clusters created and the number of nearby clusters searched can affect both the search speed and the precision of the results.

While these performance metrics provide an important indication of system scalability, the primary focus of this study was to demonstrate the system’s ability to index and search datasets that exceed available memory. It should be noted that in-memory indexing methods, such as IVF\_FLAT or HNSW, are unable to handle datasets once memory capacity is exceeded, resulting in system failures when the data surpasses the 16 GB RAM limit. In contrast, the disk-based indexing methods evaluated here functioned effectively even when the dataset size exceeded available RAM, illustrating the potential of these approaches for large-scale codebases. This behavior is reproducible under similar memory-constrained conditions and can be tested on different hardware configurations.

\section{Conclusion}
\label{conclusion}
In this paper we address the problem of searching for code clones in an industrial-size code base with more than a billion lines of code. In our specific industrial application, we have a very large \textit{corpus} of code that stays relatively stable, and we perform many \textit{queries} on the \textit{corpus} as new code is developed. Although neural networks have been applied very successfully to the pairwise \textit{clone comparison} problem, such techniques require far too many pairwise comparisons to scale to even a moderately large code base. A much more scalable approach to clone search is to create a multidimensional vector embedding for each code fragment, and find similar code fragments by searching for $k$ nearest neighbors within the multidimensional vector space. SSCD has used this approach to scale to much larger code sizes than neural network pairwise clone comparison approaches.

To address the problem of clone search at an industrial scale, we introduce DB-SSCD which extends the existing SSCD approach. We address the problem of our \textit{corpus} being extremely large in two main ways: we make it persistent on external storage so that search data structures can be reused across executions; and we make use of efficient disk-based nearest neighbour search techniques to allow the code, embeddings, and search data structures to be searched on external storage rather than being limited by in-memory techniques. In addition, we have solved specific problems that arise in industrial-scale clone search. For example, we have observed large classes of identical (or almost identical) Type 1 clones that can overwhelm $k$-nearest neighbour algorithms that rely on a relatively low value of $k$ for efficient execution.

We demonstrate that by leveraging disk-resident embeddings and advanced indexing techniques, DB-SSCD can handle industrial-scale datasets that far exceed the memory capacity of typical systems. Our solution integrates two state-of-the-art vector search libraries, Faiss and Milvus, to perform efficient nearest neighbor searches on disk. Through a comprehensive set of experiments, we have validated that:
\begin{itemize}
    \item DB-SSCD maintains high recall rates comparable to in-memory methods on standard datasets like BigCloneBench, ensuring that accuracy is not sacrificed for scalability.
    \item In the case of the BigCloneBench dataset, it is possible to compare execution time of the in-memory SSCD approach and the external storage approach of DB-SSCD. We find that DB-SSCD using a fast NVMe solid-state drive for external storage is around 2$\times$ slower than the purely in-memory approach. It is not possible to do a similar comparison for the larger BigCode ``Stack'' dataset, because the code, embeddings and search data structures are too large for the in-memory approach.
    %\item Faiss excels in indexing speed and storage efficiency, making it ideal for environments where fast indexing and low disk usage are key priorities.
    %\item Milvus achieves superior query throughput, particularly for larger, high-dimensional datasets like The Stack, making it more suitable for applications that require high search performance and throughput.
    \item Disk-based nearest neighbor search is a viable solution for industrial-scale clone detection, allowing for efficient and accurate \textit{query} search across \textit{corpus} codebases with more than a billion of lines of code.
    %\item The experiments also highlighted the trade-offs between the two systems. While Faiss provides faster indexing and is more storage-efficient, Milvus offers higher query throughput, making it ideal for real-time or high-concurrency applications. These findings demonstrate that DB-SSCD can be adapted to a variety of use cases depending on the specific needs of the software engineering environment, such as code maintenance, security audits, or large-scale refactoring.
\end{itemize}

%% Change for camera-ready version
\noindent Our work demonstrates that neural network-based clone detection methods can scale to billion-line industrial code bases. Billion-line clone detection has previously been possible, but using classical search methods such as hashing, bags of tokens, and suffix trees. But neural network clone detection has the possibility of achieving significantly higher accuracy than classical approaches. Thus, our work demonstrates the possibility of higher accuracy clone detection at billion-line scale.
%%

% In future work, we aim to incorporate the optimizations suggested in the SPANN paper to enhance the performance of Faiss based indexing and search system. These optimizations have the potential to improve the search throughput and thus  make the implementation  more competitive with Milvus based indexes. We will also evaluate our approach using an even larger corpus of source code with our industrial partners.
In future work we will evaluate our approach using an even larger corpus of source code with our industrial partners. In the current paper we search for code clones at the function/method level, where we create a vector embedding for each function/method. In future work we will also investigate clone detection at the sub-function level, which will result in multiple embeddings being stored for each function, and even larger-scale clone search challenges.

\section*{Acknowledgment}
This work was supported by Science Foundation Ireland grant 16/RC/3918 and by Huawei Technologies Co., Ltd.

\bibliographystyle{IEEEtran}
%\bibliography{refs}
\bibliography{refs}

\begin{thebibliography}{10}
\providecommand{\url}[1]{#1}
\csname url@samestyle\endcsname
\providecommand{\newblock}{\relax}
\providecommand{\bibinfo}[2]{#2}
\providecommand{\BIBentrySTDinterwordspacing}{\spaceskip=0pt\relax}
\providecommand{\BIBentryALTinterwordstretchfactor}{4}
\providecommand{\BIBentryALTinterwordspacing}{\spaceskip=\fontdimen2\font plus
\BIBentryALTinterwordstretchfactor\fontdimen3\font minus \fontdimen4\font\relax}
\providecommand{\BIBforeignlanguage}[2]{{%
\expandafter\ifx\csname l@#1\endcsname\relax
\typeout{** WARNING: IEEEtran.bst: No hyphenation pattern has been}%
\typeout{** loaded for the language `#1'. Using the pattern for}%
\typeout{** the default language instead.}%
\else
\language=\csname l@#1\endcsname
\fi
#2}}
\providecommand{\BIBdecl}{\relax}
\BIBdecl

\bibitem{Ain2019}
Q.~U. Ain, W.~H. Butt, M.~W. Anwar, F.~Azam, and B.~Maqbool, ``A systematic review on code clone detection,'' \emph{IEEE Access}, vol.~7, pp. 86\,121--86\,144, 2019.

\bibitem{Rattan2013}
D.~Rattan, R.~Bhatia, and M.~Singh, ``Software clone detection: A systematic review,'' \emph{Information and Software Technology}, vol.~55, pp. 1165--1199, 2013.

\bibitem{ahmed2024nearest}
G.~A. Ahmed, J.~V. Patten, Y.~Han, G.~Lu, W.~Hou, D.~Gregg, J.~Buckley, and M.~Chochlov, ``Nearest-neighbor, bert-based, scalable clone detection: A practical approach for large-scale industrial code bases,'' \emph{Software: Practice and Experience}, 2024.

\bibitem{Wang2021}
Y.~Wang, W.~Wang, S.~Joty, and S.~C.~H. Hoi, ``Code{T5}: Identifier-aware unified pre-trained encoder-decoder models for code understanding and generation,'' in \emph{Proceedings of the 2021 Conference on Empirical Methods in Natural Language Processing}, Sept. 2021.

\bibitem{Feng2020}
\BIBentryALTinterwordspacing
Z.~Feng, D.~Guo, D.~Tang, N.~Duan, X.~Feng, M.~Gong, L.~Shou, B.~Qin, T.~Liu, D.~Jiang, and M.~Zhou, ``Codebert: A pre-trained model for programming and natural languages,'' 2 2020. [Online]. Available: \url{http://arxiv.org/abs/2002.08155}
\BIBentrySTDinterwordspacing

\bibitem{White2016}
M.~White, M.~Tufano, C.~Vendome, and D.~Poshyvanyk, ``{Deep learning code fragments for code clone detection},'' \emph{ASE 2016 - Proceedings of the 31st IEEE/ACM International Conference on Automated Software Engineering}, pp. 87--98, 2016.

\bibitem{Wei2017}
H.~H. Wei and M.~Li, ``{Supervised deep features for Software functional clone detection by exploiting lexical and syntactical information in source code},'' \emph{IJCAI International Joint Conference on Artificial Intelligence}, pp. 3034--3040, 2017.

\bibitem{Saini2018}
V.~Saini, F.~Farmahinifarahani, Y.~Lu, P.~Baldi, and C.~Lopes, ``{Oreo: Detection of clones in the twilight zone},'' \emph{arXiv}, pp. 354--365, 2018.

\bibitem{Guo2020}
\BIBentryALTinterwordspacing
D.~Guo, S.~Ren, S.~Lu, Z.~Feng, D.~Tang, S.~Liu, L.~Zhou, N.~Duan, A.~Svyatkovskiy, S.~Fu, M.~Tufano, S.~K. Deng, C.~Clement, D.~Drain, N.~Sundaresan, J.~Yin, D.~Jiang, and M.~Zhou, ``Graphcodebert: Pre-training code representations with data flow,'' 9 2020. [Online]. Available: \url{http://arxiv.org/abs/2009.08366}
\BIBentrySTDinterwordspacing

\bibitem{Chochlov2022}
M.~Chochlov, G.~Aftab~Ahmed, J.~Vincent~Patten, G.~Lu, W.~Hou, D.~Gregg, and J.~Buckley, ``Using a nearest-neighbour, {BERT}-based approach for scalable clone detection,'' in \emph{IEEE International Conference on Software Maintenance and Evolution (ICSME)}, 2022, pp. 582--591.

\bibitem{Svajlenko2017a}
J.~Svajlenko and C.~K. Roy, ``{B}ig{C}lone{E}val: A clone detection tool evaluation framework with {B}ig{C}lone{B}ench,'' 2016, pp. 596--600.

\bibitem{Kocetkov2022TheStack}
D.~Kocetkov, R.~Li, L.~Ben~Allal, J.~Li, C.~Mou, C.~Muñoz~Ferrandis, Y.~Jernite, M.~Mitchell, S.~Hughes, T.~Wolf, D.~Bahdanau, L.~von Werra, and H.~de~Vries, ``The {S}tack: 3 {TB} of permissively licensed source code,'' \emph{arXiv:2211.15533}, 2022.

\bibitem{Kamiya2002654}
T.~Kamiya, S.~Kusumoto, and K.~Inoue, ``Ccfinder: A multilinguistic token-based code clone detection system for large scale source code,'' \emph{IEEE Transactions on Software Engineering}, vol.~28, pp. 654--670, 2002.

\bibitem{Jiang2007}
L.~Jiang, G.~Misherghi, Z.~Su, and S.~Glondu, ``Deckard: Scalable and accurate tree-based detection of code clones,'' 2007, pp. 96--105.

\bibitem{Malkov2020}
Y.~Malkov and D.~Yashunin, ``Efficient and robust approximate nearest neighbor search using hierarchical navigable small world graphs,'' \emph{IEEE Trans. Pattern Anal. Mach. Intell.}, vol.~42, no.~4, p. 824–836, April 2020.

\bibitem{johnson2019billion}
J.~Johnson, M.~Douze, and H.~J{\'e}gou, ``Billion-scale similarity search with {GPU}s,'' \emph{IEEE Transactions on Big Data}, vol.~7, no.~3, pp. 535--547, 2019.

\bibitem{Ranzato_NEURIPS2021}
Q.~Chen, B.~Zhao, H.~Wang, M.~Li, C.~Liu, Z.~Li, M.~Yang, and J.~Wang, ``{SPANN}: Highly-efficient billion-scale approximate nearest neighborhood search,'' in \emph{Advances in Neural Information Processing Systems}, vol.~34.\hskip 1em plus 0.5em minus 0.4em\relax Curran Associates, Inc., 2021, pp. 5199--5212.

\bibitem{Jayaram_NEURIPS2019}
S.~Jayaram~Subramanya, F.~Devvrit, H.~V. Simhadri, R.~Krishnawamy, and R.~Kadekodi, ``Disk{ANN}: Fast accurate billion-point nearest neighbor search on a single node,'' in \emph{Advances in Neural Information Processing Systems}, vol.~32.\hskip 1em plus 0.5em minus 0.4em\relax Curran Associates, Inc., 2019.

\bibitem{Wang_Milvus_2021}
J.~Wang, X.~Yi, R.~Guo, H.~Jin, P.~Xu, S.~Li, X.~Wang, X.~Guo, C.~Li, X.~Xu, K.~Yu, Y.~Yuan, Y.~Zou, J.~Long, Y.~Cai, Z.~Li, Z.~Zhang, Y.~Mo, J.~Gu, R.~Jiang, Y.~Wei, and C.~Xie, ``Milvus: A purpose-built vector data management system,'' in \emph{Proceedings of the 2021 International Conference on Management of Data}, ser. SIGMOD '21.\hskip 1em plus 0.5em minus 0.4em\relax New York, NY, USA: Association for Computing Machinery, 2021, p. 2614–2627.

\bibitem{jegou2010product}
H.~Jegou, M.~Douze, and C.~Schmid, ``Product quantization for nearest neighbor search,'' \emph{IEEE transactions on pattern analysis and machine intelligence}, vol.~33, no.~1, pp. 117--128, 2010.

\bibitem{Mou2016}
L.~Mou, G.~Li, L.~Zhang, T.~Wang, and Z.~Jin, ``Convolutional neural networks over tree structures for programming language processing,'' in \emph{Proceedings of the Thirtieth AAAI Conference on Artificial Intelligence}, ser. AAAI'16.\hskip 1em plus 0.5em minus 0.4em\relax AAAI Press, 2016, p. 1287–1293.

\bibitem{Sajnani2016}
H.~Sajnani, V.~Saini, J.~Svajlenko, C.~K. Roy, and C.~V. Lopes, ``Sourcerercc: Scaling code clone detection to big-code,'' 2016, pp. 1157--1168.

\bibitem{Li2020}
G.~Li, Y.~Wu, C.~K. Roy, J.~Sun, X.~Peng, N.~Zhan, B.~Hu, and J.~Ma, ``Saga: Efficient and large-scale detection of near-miss clones with gpu acceleration.''\hskip 1em plus 0.5em minus 0.4em\relax IEEE, 2020, pp. 272--283.

\end{thebibliography}
\end{document}